\RequirePackage{etoolbox}
\csdef{input@path}{{style/}{graphics/}}
\documentclass[MSNbibl,nameyear,dvips]{arxstspdf}
\usepackage{url,breakurl}
\usepackage{flushend}
\usepackage{stfloats}
\usepackage{graphicx}

\volume{30}
\issue{1}
\pubyear{2015}
\firstpage{59}
\lastpage{71}
\doi{10.1214/14-STS484} % kopijuoti is 'New paper accepted'
\makeatletter
\newcommand{\rrvert}{\vert}
\newcommand{\llvert}{\vert}
\renewcommand{\citep}[1]{(\cite{#1})}

\newcommand{\Var}{\operatorname{Var}}
\newcommand{\Cov}{\operatorname{Cov}}
\newcommand{\Binomial}{\operatorname{Binomial}}

\newcommand{\E}{\mathrm{E}}
\newcommand{\Prob}{\mathrm{P}}

\makeatother

\begin{document}
\begin{frontmatter}

\title{Monte Carlo Null Models for~Genomic~Data}%\thanksref{T1}
% kai straipsnis turi susijusiu diskusiju ir rejoinder'iu
%rejoinder at \relateddoi{r}{10.1214/00-STSXXXX}.}
\runtitle{Monte Carlo null models}

\begin{aug}
\author[A]{\fnms{Egil}~\snm{Ferkingstad}\corref{}\ead[label=e1]{egil@hi.is}},
\author[B]{\fnms{Lars}~\snm{Holden}\ead[label=e2]{Lars.Holden@nr.no}}
\and
\author[C]{\fnms{Geir Kjetil}~\snm{Sandve}\ead[label=e3]{geirksa@ifi.uio.no}}
\runauthor{E. Ferkingstad, L. Holden and G. K. Sandve}

\affiliation{University of Iceland and Norwegian Computing Center,
Norwegian Computing Center and University of Oslo}

\address[A]{Egil Ferkingstad is Research Scholar,
Science Institute, University of Iceland, Dunhaga 5, 107 Reykjavik,
Iceland, and
Norwegian Computing Center,
Gaustadalleen 23B,
0373 Oslo, Norway \printead{e1}.}
\address[B]{Lars Holden is Managing Director, Norwegian Computing Center,
Gaustadalleen 23B,
0373 Oslo, Norway \printead{e2}.}
\address[C]{Geir Kjetil Sandve is Associate Professor,
Department of Informatics,
University of Oslo,
Gaustadalleen 23B,
0373 Oslo, Norway \printead{e3}.}
\end{aug}

% ABSTRACT
%
\begin{abstract}
As increasingly complex hypothesis-testing scenarios are considered
in many scientific fields, analytic derivation of null distributions
is often out of reach. To the rescue comes Monte Carlo testing,
which may appear deceptively simple: as long as you can sample test
statistics under the null hypothesis, the $p$-value is just the
proportion of sampled test statistics that exceed the observed test
statistic. Sampling test statistics is often simple once you have a
Monte Carlo null model for your data, and defining some form of
randomization procedure is also, in many cases, relatively
straightforward. However, there may be several possible choices of
a randomization null model for the data and no clear-cut criteria for choosing
among them. Obviously, different null models may lead to very
different $p$-values, and a very low $p$-value may thus occur
due to the inadequacy of the chosen null model. It is
preferable to use assumptions about the underlying random data
generation process to guide selection of a null model.
In many cases, we may order the null models by increasing
preservation of the data characteristics, and we argue in this paper
that this ordering in most cases gives increasing \mbox{$p$-values}, that is,
lower significance. We
denote this as \textit{the null complexity principle}. The principle
gives a better understanding of the different null models
and may guide in the choice between the different models.
\end{abstract}

% KEYWORDS
\end{frontmatter}

%s1 #&#
\section{Introduction}\label{sec:introduction}

Increasingly, Monte Carlo methods are needed to provide answers to
important scientific questions, particularly in the rapidly advancing
field of genomics. For better or worse, these questions are often
framed within the formalism of statistical hypothesis testing. In many
cases, Monte Carlo hypothesis testing techniques such as permutation
testing are the only options.

Conceptually, these methods share an appealing clarity: As long as you
can sample test statistics under the null hypothesis, the $p$-value is
just the proportion of sampled test statistics that exceed the
observed test statistic.
One of our main aims is to show that the apparent simplicity of
randomization hypothesis testing can be very deceptive. In the
following, we use \emph{null model} as a general term for the
distribution of the resampled data (e.g., using random permutations),
and we use \emph{null distribution} to denote the distribution of the
test statistic under the null model. Even though there is a highly
developed theory of classical hypothesis testing (e.g., \cite{lehmann2005}),
new practical and methodological problems
appear when we need to resort to Monte Carlo testing:
\begin{itemize}
\item The question of interest may be unavoidably vague, so that it is
not obvious how to translate it into a precise mathematical
formulation.
\item There may be several possible choices of a randomization null
model and no clear-cut criteria for choosing among them (except
possibly conservativeness arguments for choosing the null model
giving the largest $p$-values).
\item A full specification of the null hypothesis consists of both
the null model and the question of interest. This complicates the
interpretation of a rejection of the null hypothesis---the
question of interest may not really have been answered if the null
model is inadequate.
\item There may be several possible choices of test statistic and
no clear-cut criteria for choosing one (except possibly
power considerations).
\end{itemize}

If unresolved, these problems may degrade the reproducibility and
transparency of investigations, as well as lead to false research
findings. There has lately been an increasing focus on how to make
science more reproducible, especially in the field of computational
biology (\cite{Ioannidis:2008cr}; \cite{noseda08where};
\cite{mesirov10accessible}; \cite{sandve2013tenrules}). Also, due to the increased
prevalence of
data-driven science \citep{douglas04here} through increased
availability of public data and more accessible and efficient
analytical tools, there has also been a heated discussion on
whether a large proportion of published research findings are false
(\cite{ioannidis2005}; \cite{goodman2007}). We discuss this topic further
in the remainder of this paper.
Our main application of interest is genomics and the Genomic
HyperBrowser (\citeauthor{sandve2010}, \citeyear{sandve2010}, \citeyear{sandve2013}) where choosing the correct
null model
is a major issue. We have discussed null models in ecology in a
companion report, \citet{ferkingstad2013}.
Several examples show that the choice of a null model can strongly
affect the resulting $p$-values. We state that ordering the null models
according to increasing preservation may imply an ordering of the statistical
significance. Further, if the null models are not able to capture the
essential structural properties of data, this may lead to false
findings.

We proceed as follows: Section~\ref{sec:randomization} discusses
general problems of randomization null models.
Section~\ref{sec:ordering} presents null model preservation
hierarchies and significance orderings.
Sections~\ref{sec:genomiclocation}--\ref{sec:Falserejections}
illustrate
several different null models within genomics:
Section~\ref{sec:genomiclocation}
considers null models for the location of transcription factor
binding sites, Section~\ref{sec:chromatinStates} shows that genetic
properties have a tendency to cluster along the genome, while Section~\ref{sec:Falserejections} illustrates that we may
get false rejections with too simple null models using simulated data of
points and segments in genomic tracks.
Finally,
Section~\ref{sec:discussion} provides a general discussion and some
concluding remarks and recommendations.

For the genomics case studies described in
Sections~\ref{sec:genomiclocation}--\ref{sec:Falserejections}
we have
used $q$-values \citep{storey2002} to correct for multiple testing.
Assume that we test $m$ hypotheses where
$p_{(1)} \leq p_{(2)} \leq\cdots\leq p_{(m)}$ are the ordered,
observed $p$-values, $R$ is the number of rejected null hypotheses, and
$V$ is the (unknown) number of falsely rejected null hypotheses. The
false discovery rate (FDR) \citep{benjamini1995} is then defined as
FDR $=\E(V/R)$. For each test,
the corresponding $q$-value is defined as the minimum FDR at which the
test is called
significant. Let $\pi_0$ be the proportion of tests that are truly null
\citep{langaas2005} and $q_{(i)}$ the $q$-value for the test with
$p$-value $p_{i}$. Then, we may estimate $q_{(i)}$ by
\[
\hat q_{(i)} = \min_{i \leq j \leq m} m * \hat\pi_0 *
p_{(j)} / j,
\]
where $\hat\pi_0$ is an estimate of $\pi_0$.
Thus, the main inputs to this multiple testing method are
the observed $p$-values together with an estimate of $\pi_0$. To
estimate $\pi_0$, we have used the robust estimator
of \citet{pounds2006}, since this is very computationally efficient
and can be shown to be conservative in many realistic settings. For
a general discussion of multiple-testing issues in Monte Carlo
settings, see also \citet{sandve2011}. All calculations were
performed using the R programming language \citep{R} and the
Genomic HyperBrowser. A Galaxy Pages \citep{goecks2010} document
allowing for replication of the results is available at
\url{https://hyperbrowser.uio.no/suppnullmodels}.

%s2 #&#
\section{Randomization Null Models}\label{sec:randomization}

Consider a hypothesis test based on data $X$ and a test statistic $T=T(X)$.
Without loss of
generality, we may assume that large values of $T$ constitute evidence
against $H_0$. Then, for an observed test statistic $T=t$, the
decision to accept or reject $H_0$ can be based on the $p$-value
$p=F_0(T \geq t)$, where we reject $H_0$ if $p<\alpha$ for some
threshold $\alpha$ and where $F_0$ is the distribution of $T$ under
the null model $P_0$. If $P_0$ is false, $T$ has distribution $F_1$.

In the classic textbook setting, the null model is known and can be
described explicitly, so we can directly compute the
$p$-value. Increasingly, both data and models are too complex for this
to be done. In such cases we must resort to some type of
Monte Carlo randomization test: we generate samples $T_i=t_i$,
$i=1,\ldots,n$ of the test statistic $T$ under the null model and
estimate the empirical $p$-value from the data set $X$ by
%
%e1 #&#
\begin{equation}\label{p-val2}
\hat{p}_{X,e}(t) = \frac{1}{n} \sum_{i=1}^{n}
I(t_i \geq t),
\end{equation}
where $t$ is the observed test statistic and $I(\cdot)$ denotes the
indicator function, equal to one if its argument is true or zero
if false. The idea of randomization testing has been around at least
since the pioneering work of \citet{fisher1935}, but has only
become practical with the advent of electronic computers. For a recent
overview of Monte Carlo methods, see \citet{manly2007}.

The randomization null model is arguably the most crucial component of
the Monte Carlo testing setup. Often, the research question and even
the test statistics may be clear, but how should one specify the null
model?
\citet{sandve2010} introduce the idea of
null model preservation hierarchies and note that ``a crucial aspect
of an investigation is the precise formalization of the null model,
which should reflect the combination of stochastic and selective
events that constitutes the evolution behind the observed genomic
feature. [\ldots] Unrealistically simple null models may [\ldots] lead to
false positives.'' Here, we build further on these ideas and provide
a conceptual framework to aid the choice of null model.

In the statistics literature, the most directly relevant previous papers
on null models are \citet{efron2004} and
\citet{bickel2010}. \citet{efron2004} estimates the null model from
data in multiple-testing problems, giving an ``empirical null.'' This
is very useful for some multiple-testing settings, but not directly
applicable to the problems we study here. \citet{bickel2010} propose
subsampling methods based on a piecewise stationary model for genome
sequences, a potentially useful approach for our case
study in Section~\ref{sec:genomiclocation}, but which we feel
would be beyond the scope of this paper.

There is also relevant work from other
disciplines. Particularly, null models have been a very contested
issue within ecology, as further discussed in \citet{ferkingstad2013}.
For example, \citet{gotelli2000} points out
that ``the analysis of presence--absence matrices with null model
randomization tests has been a major source of controversy in
community ecology for over two decades.'' See also the book
by \citet{gotelli1996} and \citeauthor{manly2007} [(\citeyear{manly2007}), Chapter~14], who notes
that ``one of the interesting aspects of this [species competition
problem] is the difficulty in defining the appropriate model of
randomness'' (page 348). \citet{fortin2000} discuss randomization tests
for spatially autocorrelated data. As discussed elsewhere in this
paper, genomics is another
area where the problem of choosing the right null model is very
urgent \citep{sandve2010}. \citet{bickel2010} note that ``a common
question asked in many applications is the following: Given the
position vectors of two features in the genome [\ldots] and a measure of
relatedness between features [\ldots] how significant is the observed
value of the measure? How does it compare with that which might be
observed `at random?' The essential challenge in the statistical
formulation of this problem is the appropriate modelling of randomness
of the genome, since we observe only one of the multitudes of possible
genomes that evolution might have produced for our and other
species.'' See \citet{kallio2011} for a general discussion of the
importance of null models within bioinformatics. Related work has also
been done within the field of data mining;
see \citet{gionis2007}, \citet{hanhijarvi2009}. \citet{lijffijt2010} consider
the related problem of estimating the level of preservation needed to
attain a prespecified significance level $\alpha$ (for example,
$\alpha=0.05$).

%s3 #&#
\section{Preservation and Significance~Orderings}\label{sec:ordering}

By assumption, the data set $X$ is taken as given, that is, it is not
considered to be a random sample from some population. In order to test
our hypothesis, we need to randomize $X$ from a null model $P_0$.
In many cases some specific features of $X$ will need to be preserved.
In a
specific problem, it may be very difficult to decide what features are
fundamental and which are not. If we attempt to conserve all possible
features of the observed $X$, we are left with $X$ itself and no
basis for performing the hypothesis test. If we conserve too little,
we generate realizations that
violate basic properties of the phenomenon under study. Different null
models may preserve different properties of $X$, for example, null model
$P_0$ preserves properties Q and R and null model $P_1$ preserves properties
R and S. But quite often we
may order the null models according to increasing preservation of the
properties of $X$. We describe two
different alternative descriptions
of ordering of preservation of the null models:
\begin{enumerate}
\item[A.] Let $P_0$ denote the state space obtained by a set of resamplings
(for example, permutations) that are allowed under a given null
model. That is, the state space is the set of all possible combinations
of values of variables in the stochastic model. %%Then the null model
%is finite with equally likely states.
We
define a preservation hierarchy
if the following criteria are satisfied: $P_0^{(1)} \subset
P_0^{(2)}\subset\cdots\subset P_0^{(n)}$. We then state that
$P_0^{(i)}$ preserves more than $P_0^{(i+1)}$ for $i=1,2,\ldots, n-1$
of the properties of the original data set $X$ and hence is more
restricted. As we will discuss further below, a more restricted null
model will in most cases give less significant results, that is,
$p$-values from $P_0^{(i)}$ will tend to be larger than $p$-values from
$P_0^{(j)}$ if $P_0^{(i)} \subset P_0^{(j)}$. Note that we only
consider \textit{Monte Carlo} null models, that is, null models that
are generated by resampling from the observed data (as in permutation
testing), and that the $P_0^{(i)}$ are sets of allowed resamplings
under $H_0$---they are \textit{not} sets of allowed parameter values.

\item[B.] Let $X=(X_1,X_2,\ldots,X_n)$ denote a state in the state
space and
let the null model be defined by a
set of allowed
permutations of the $X_i$'s. Define $X_i=1$ for a certain property in
base pair $i$ and otherwise
$X_i=0$. Assume further that the test statistic $T$ is given by
%
%e2 #&#
\begin{equation}
\label{teststat-sum} T=\frac{1}{n}\sum_{i}
y_i X_i
\end{equation}
%
% = \frac{1}{n}\sum_{i} Z_i\]
for a fixed vector $y=(y_1,y_2,\ldots,y_n)$. We trivially have
\[
\E(T)=\frac{1}{n}\sum_{i} y_i
\E(X_i)
\]
and
\[
\Var(T)=\frac{1}{n^2}\sum_i \sum
_j y_i y_j \Cov(X_i,X_j).
\]
%
% = \frac{1}{n^2}\sum_i \sum_j \Cov(Z_i,Z_j). \]
We assume the stationary criteria $\E(X_i)=\lambda$ and~$\Var(X_i)=\sigma^2$ are independent of $i$.
Assume $\Cov(X_i,X_j)$ is positive for $\llvert i-j\rrvert $ small and decreases with
increasing distance $\llvert i-j\rrvert $, say, $\Cov(X_i,X_j)=\sigma^2 \rho(\llvert i-j\rrvert )$,
for some decreasing, positive
correlation function $\rho$. The covariance is smaller in null model
$P_{(1)}$ than $P_{(2)}$ if
the corresponding correlation functions satisfy
$\rho^{(1)}(d) \geq\rho^{(2)}(d)$ for all $d>0$.
This implies that the more the permutation preserves
of $\Cov(X_i,X_j)$ for $\llvert i-j\rrvert $ small, the larger is $\Var(T)$.
Here we may define a sequence of null models with
decreasing $\Cov(X_i,X_j)$ for all distances $\llvert i-j\rrvert $, implying larger
values of
$\Var(T)$. In most cases it is reasonable to also assume that $\E(T)$ is
the same for all the null models.
\end{enumerate}

Cases A and B may both be satisfied at the same time.
In Section~\ref{sec:chromatinStates} we argue that it is typical for
genomic data of certain types to satisfy
the criteria in case B, that is, $\Cov(X_i,X_j)$ is positive for
$\llvert i-j\rrvert $ small and decreases with
increasing distance $\llvert i-j\rrvert $.
In this case, we
make assumptions directly on the test statistic $T$ which indicate
larger empirical $p$-values [see definition (\ref{p-val2})] the more we
preserve of the original data~$X$.
By assumption, large values of $T$ indicate evidence against the
hypothesis $H_0$.
%we have observed the value $t > \E(T)$.
A larger value of $\Var(T)$ implies under quite general
statistical assumptions that a larger fraction of the realizations
have a test statistic $T_i>T$ (provided the number of realizations are
sufficiently large), leading to larger $p$-values. Also, in case A,
an increasing state space will in most cases lead to an
increase in $\Var(T)$.

The relationship between preservation and significance
is the same observation as in \citet{hanhijarvi2009}, ``obviously,
the more restricted the null hypothesis [\ldots] the less significant the
results of a data mining algorithm tend to be.''
We will call this observation the \emph{null complexity principle}.

The null complexity principle may be an aid in choosing the correct
level of preservation in the null model, as well as in interpretation
of the results. Since the null complexity principle does not always
hold, it is necessary to demonstrate it for the problem under
study. If this property is proved for the null models applied, then
this is very useful information when choosing a null model. For
example, a scientist wishing to be conservative may choose the null
model known {a priori} to give the largest $p$-values. Also,
some Monte Carlo null models may be considerably more computationally
demanding than others. Then, we may first test a null model having low
computational cost. If we reject the hypothesis using this model, we
will also reject the hypothesis for less conservative (and more
computationally intensive) null models. The ordering of the $p$-values
imply that too simple null models may lead to false positives, as
conjectured in \citet{sandve2010}.

Our concepts of null models and preservation may be illustrated by the
following simple example. Assume we have tossed a coin $N\gg 100$ times
and we question whether the observed proportion of heads in the
beginning of the sequence is significantly larger than $0.5$. We want
to allow for the possibility of coins tosses being correlated. We use
the number of heads in the first $100$ coin tosses as the test
statistic. We use two different null models. In null model 1 we assume
that the coins are independent of each other and have a 50\%
probability for heads, so we can permute the observed coin tosses
freely to sample from the null model. For null model 2 we permute each
sequence of 2 observations from the observed $N$ coins in order to
maintain a possible correlation between consecutive coins. The second
model is more restrictive and according to the null complexity
principle gives larger $p$-values. If there is positive correlation
between consecutive coins, this increases the variability of the test
statistics and hence increases the $p$-value. However, if there is
negative correlation between consecutive coins, this decreases the
variability of the test statistics and hence decreases the $p$-value.
The example also illustrates that the null complexity principle often
assumes positive correlations between terms in the test statistic.
For test statistics defined on point processes (such as the examples in
Section~\ref{sec:genomiclocation}), this typically corresponds to
attraction between points (correlation between consecutive inter-point
distances). Intuitively, it is easier to envision mechanisms leading to
attraction than repulsion (although these for sure also exist). Our
experience is that positive correlations (including attraction in
point processes) are much more common than negative correlations
(including repulsion) in real data sets, which we also show for a
number of genomic data sets, representing several classes of features,
in Section~\ref{sec:chromatinStates}.

%s3.1 #&#
\subsection{How to Measure Clustering of Points}\label{sec:clustering}

As we have seen in case B above, in some cases it is important to preserve
clustering of points, since this has important implications for the
sizes of the resulting $p$-values.
Following the notation defined in the previous section, we may use the
Ripley's $K$-function \citep{ripley1976} as a measure for clustering.
This is defined relative
to a distance $t$ as
\begin{eqnarray*}
&&\hspace*{-7pt}K(t)=\lambda^{-1}\E(\mbox{number of extra points within}% distance } t
\\
&& \hspace*{56pt}\mbox{distance } t \mbox{ of a randomly chosen point}).
\end{eqnarray*}
To simplify the notation, disregard edge effects by assuming that there
exist $X_{-t-1},\ldots,X_0$ and
$X_{n+1},\ldots,\allowbreak X_{n+t}$ from the same process as $X_1,\ldots,X_n$. Then
\[
K(t) = (n\lambda)^{-1} \sum_{i=1}^n
\mathop{\sum_{j=i-t}}\limits
_{j
\neq i}^{i+t}
\Prob(X_j=1|X_i=1)
\]
for integer $t$.
We may write $K(t)$ in terms of the correlation function $\rho$, as follows:
\begin{eqnarray*}
K(t) &=& (n\lambda)^{-1} \sum_{i=1}^n
\mathop{\sum_{j=i-t}}\limits
_{j \neq i}^{i+t} \bigl(\lambda+
\lambda ^{-1}\Cov(X_i,X_j) \bigr)
\\
&=& 2t + \sigma^2 n^{-1}\lambda^{-2}\sum
_{i=1}^n \mathop{\sum
_{j=i-t}}\limits
_{ j \neq i}^{i+t} \rho\bigl(\llvert i-j\rrvert
\bigr)
\\
&=& 2t + 2 \sigma^2 n^{-1} \lambda^{-2} \sum
_{i=1}^n \sum_{j=1}^{t}
\rho (j)
\\
&=& 2t + 2 \sigma^2 \lambda^{-2} \sum
_{j=1}^{t} \rho(j).
\end{eqnarray*}
Using our earlier definition of clustering [$\rho^{(1)}(d) \geq\rho
^{(2)}(d)$ for all $d>0$], this means that increased clustering implies
increased $K(t)$ for each $t$.

Note that if $X_i$ and $X_j$ are independent for $i \neq j$, then
\[
K(t) = (n\lambda)^{-1}\sum_{i=1}^n
\mathop{\sum_{j=i-t }}\limits
_{j
\neq i}^{i+t} \lambda = (n
\lambda)^{-1}\sum_{i=1}^n (2t
\lambda ) = 2t.
\]
Therefore, we may define a scaled $K$-function, $L(t)$, as follows:
\[
L(t)=K(t)/(2t).
\]
Then, $L(t)<1$ corresponds to repulsion between points, $L(t)=1$ to
independent points,
while $L(t)>1$ corresponds to attraction between points.

Assume that we have observed $X_i=x_i$, $i=1,\ldots,n$ and wish to
estimate $\hat L(t)$. To simplify notation,
let $x_i=0$ for $i<1$ and $i>n$.
Then, we choose some value $t=\tau$ and estimate $K(\tau)$ by
\[
\hat K(\tau) = n^{-1}\hat\lambda^{-2}\sum
_{i=1}^n \mathop{\sum_{j=i-\tau}}_{ j \neq i}^{i+\tau}
w_{ij}^{-1} x_i x_j,
\]
where
\[
\hat\lambda=n^{-1}\sum_{i=1}^n
x_i
\]
and
\[
w_{ij}=\frac{\min(\max(i,j),n)-\max(\min(i,j),1)}{\max(i,j)-\min(i,j)}
\]
are weights that correct for edge effects. Finally, $L(\tau)$ is
estimated by
\[
\hat L(\tau)=\hat K(\tau)/(2\tau).
\]

%s4 #&#
\section{Null Models for Genomic Locations}\label{sec:genomiclocation}
In this section we will show how to choose a null model when we want to test
whether the points in a point track are independent of segments in a
segment track. Several null models that have preservation
orderings according to both cases A and B in Section~\ref{sec:ordering}
are presented. The
results are as expected, with more preservation giving larger $p$-values.

A fully extended human chromosome would be about one meter long,
consisting of about 3 billion base pairs. The properties vary along the
genome and we often divide the genome into bins and perform separate
tests for each bin. There are about 30,000 genes,
represented as intervals of base pairs or segments in the terminology
of \citet{sandve2010}. Transcription factors (TF) regulate the
expression of genes by binding to DNA in the spatial proximity of the
genes they regulate, interacting with the complex of proteins that
transcribes DNA to RNA (the transcriptional machinery). As the DNA
may form loops, spatial proximity is not necessarily the same as
proximity along the sequence. A TF that binds to DNA may therefore
regulate the expression of a gene that is millions of base pairs away
from the binding site, and may even regulate genes on different
chromosomes (\cite{visel09genomic}; \cite{ruf11large-scale}). In higher
organisms, such as humans, transcription factor binding sites are
organized into modular units, often referred to as cis-regulatory
modules (CRM). These CRM usually comprise a few hundred base pairs
and are characterized by a high local frequency of binding for one or
several TFs (\cite{berman02exploiting}; \cite{zhou04cismodule}). TFs that
interact with the transcriptional machinery to increase the expression
of genes at some distance from where the TFs bind to DNA are often
referred to as enhancers, and the regions of DNA containing such TF
binding sites are often referred to as enhancer regions. The TF are also
segments of base pairs, but since these segments usually are shorter
than the genes, these are often represented as unmarked points in the
terminology of \citet{sandve2010}.

%s4.1 #&#
\subsection{Specifying Details of Hypothesis Tests: Transcription
Factor Binding Relative to Genes}\label{sec:transgenes}
In this section we will discuss two null models that have a preservation
ordering according to both cases A and B of Section~\ref{sec:ordering}. The
results are as expected: more preservation gives larger $p$-values.
We only get rejection of the null hypothesis when we have
little preservation. This may be due to a too simple null model.

A very basic question related to the positioning of transcription
factor binding sites (TFBS) is whether the binding sites of a given TF
fall preferentially inside or outside genes. As a concrete example,
we consider binding sites for the transcription factor MitF
\citep{strub11essential} in relation to Ensembl gene
regions \citep{Flicek:2012ks}. We asked this question locally along the
genome, dividing the genome into bins and performing one separate
test per bin. As bins we used chromosome bands, which represent a
common partition of chromosomes into regions of a few megabases. To
ensure a reasonable amount of data for the tests, we only considered
chromosome bands containing at least one gene and five TFBS, resulting
in 73 bins. Separate tests were performed for each bin.

How can a hypothesis test be specified for this problem? Clearly, a
natural test statistic is the number $T$ of TFBS falling inside
genes. Furthermore, let $n$ be the total number of TFBS in the bin and
$p$ the proportion of the bin covered by genes. A natural null
model is that TFBS are uniformly and independently located within each
bin. It is then easily seen that the distribution of the test statistic
is $T \sim
\Binomial(n,p)$. There are other alternatives. For instance,
one might assume that the TFBS are Poisson distributed within the
bin. This would preserve the underlying probability of observing a
TFBS instead of the exact count of observed TFBS, thus giving rise to
a (slightly) different null distribution. In our opinion, when
realizations are based on Monte Carlo analysis, it is necessary to
carefully study the properties of the null model. Mistakes are easily made
if one directly writes down the null distribution of the test
statistic.

Performing the binomial test as described above yields the conclusion
that there is preferential location inside genes for 9 out of the 73
bins after
multiple testing correction (at a 10\% false discovery rate).
This could be taken as an indication of local
variation of an underlying (mechanistic) tendency of  TFBS for the
transcription factor MitF to be located inside gene regions.

\begin{figure}

\includegraphics{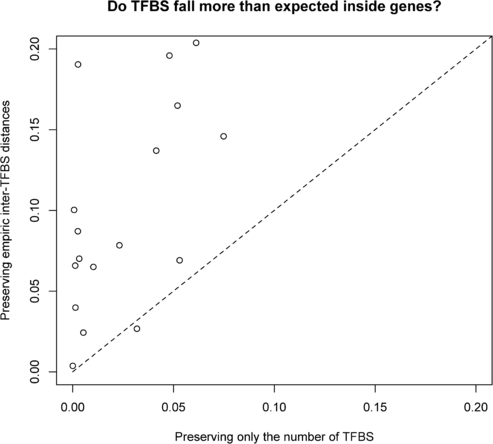}

\caption{Scatter plot of $p$-values for the same test under two
different null models.}\label{fig:uniformVsInterPoint}
\end{figure}

The TFBS may form clusters, denoted CRM, with typical length of a
few hundred base pairs. This is a much smaller scale than the gene
regions, which typically are several thousand base pairs. The
clustering of TFBS appears to be an intrinsic property of the TFBS
themselves, and not a part of the TFBS--gene relation that is being
tested. This suggests that at least some aspects of clustering should
be preserved in the null model.
This is an example of case B of Section \ref{sec:ordering}, as can be seen
by letting $X_i=1$ for a TFBS in base pair $i$. Most of the clusters are either completely
inside or completely outside a segment, meaning that
$\Cov(X_i,X_j)$ is larger for $i$ and $j$ close. If we
maintain this positive correlation in the null model, this gives
higher $p$-values. This is tested by using two different null models.
The first model is the null model described above, where we only
preserve the total number of TFBS. In the second model the empirical
inter-TFBS distances are preserved in the null model by only permuting
these distances. This second model preserves more of positive
correlation in $\Cov(X_i,X_j)$.
These two null models are in
fact also an example of case A in Section~\ref{sec:ordering}, since
both null models give a finite state space with equally likely states
and the second null model is a subset of the first one. The $p$-values
from the two null models are illustrated in
Figure~\ref{fig:uniformVsInterPoint}. We see clearly that preserving
the empirical inter-TFBS distances in the null model gives larger
$p$-values. Some bins show very different results between null models,
for example, at chromosome band q25.1, where independent location
gives a $p$-value less than 0.0005, while preservation of inter-TFBS
distance gives a $p$-value of 0.1. This is probably due to strong
correlation $\Cov(X_i,X_j)$ in this bin. When the empirical
distribution of inter-TFBS distances is preserved, the null
hypothesis is not rejected in any bin at 10\% FDR, suggesting that the
significant findings under the uniformity assumption may simply be due
to inadequacy of the null model.

%f1 #&#

%s4.2 #&#
\subsection{Deciding What Should Be Preserved in the Null Model:
Randomizing Genes Instead of Transcription Factors Binding Sites}\label{sec:randomgenes}

In this section there are
two pairs of null models with preservation ordering according to case A
in Section~\ref{sec:ordering}. The $p$-values are ordered as expected:
more preservation gives larger $p$-values.

%f2 #&#
\begin{figure}[b]

\includegraphics{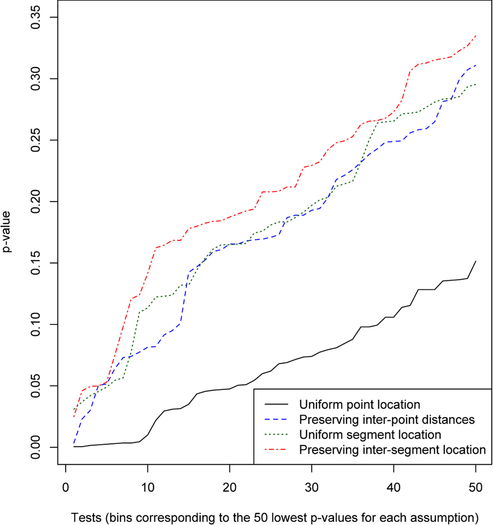}

\caption{Empirical cumulative distribution of $p$-values under four
different null models. The different null models correspond to whether TFBS
or genes are randomized, and whether the empirical inter-element
distances are preserved or not.}\label{fig:lowest50pvalues}
\end{figure}

In the above discussion, we have implicitly assumed that the TFBS
distribution should be stochastic in the null model, while genes are preserved
exactly at their genomic locations. This seems reasonable from a
biological standpoint, as the location of binding sites can generally
be assumed to follow the location of genes chronologically through
evolution (although there may be exceptions, such as coding regions
copied into genomic regions that already have an established regulatory
machinery). However, one should also consider
which of the tracks have the more complex
structure. This structure should be preserved in the null model, and
one would prefer to randomize the track with the simplest
structure. Although the location of genes is clearly not uniform, it
can be argued that the TFBS has an even more complex structure. The
reason is that individual TFBS fall as clusters with specific
intra-cluster structure inside regulatory regions, with regulatory
regions again having a certain structure in relation to genes. Indeed,
as can be seen from Figure~\ref{fig:lowest50pvalues}, the $p$-values
are somewhat higher when randomizing genes as opposed to TFBS.
In the figure we compare the two null models described above
randomizing TFBS-positions and two null models where we randomize the gene
locations with random positioning and preserving inter-gene distances.
These two null models, randomizing the gene locations, are also
examples of case A in Section~\ref{sec:ordering}. As expected, the
second null model gives larger $p$-values. The two models with gene
randomization
also give larger $p$-values than the two models with TFBS randomization,
indicating that the models with gene randomization preserve more of the
complex interaction between genes and TFBS than the two other models.
Note also that the difference between the two models randomizing genes
is smaller than between the two models randomizing TFBS. This indicates
that preserving inter-distances is more important for TFBS than for genes.

%s5 #&#
\section{Significance Ordering for Data That~Display Internal
Clustering: Transcription Factor Binding and Chromatin States}\label{sec:chromatinStates}

In this section we will show that clustering is present in a
large amount of genomic tracks. Clustering leads to the preservation ordering
shown in case B of Section~\ref{sec:ordering}. Again, the $p$-values
are ordered, with
more preservation giving larger $p$-values.

%f3 #&#
\begin{figure*}[b]

\includegraphics{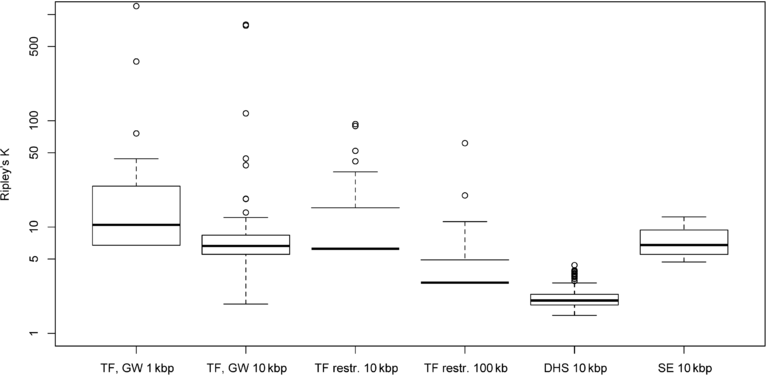}

\caption{Box plot of scaled Ripley\emph{'}s $K$ values for several
collections of ENCODE and RoadMap Epigenomics tracks.
The two left boxes are based on 81 TF
ChIP-seq tracks with genome wide data, followed by two boxes with
the same data but restricted to selected regions of size~$>100$~kb
that are more than 100~kbps away from the nearest gene, followed
by a box based on 147 tracks of DHS for different cell types and
finally a box with elements of chromatin state \emph{``}5-Strong
Enhancer\emph{''} in nine different cell types. The clustering is analyzed for
two different scales for the two
first data types.}\label{fig:Kboxplot}
\end{figure*}

The DNA has to be highly compacted in order to fit into a cell. At the
same time, it has to be accessible, for example, to the binding of
transcription factors in order to allow efficient gene regulation. To
achieve controlled compactness and accessibility, DNA is packed in a
structured manner at multiple levels. The first such organizational
layer consists of the DNA double helix, at the order of 100 base
pairs, wound around small protein complexes called nucleosomes
\citep{Kornberg99twenty-five}. These nucleosomes can be modified
through the attachment of other molecules to the proteins of the
nucleosomes, which are called histones. This is referred to as histone
modification, and serves a regulatory role in itself
\citep{cairns09logic}. Recently, it has become possible to create
genome-wide maps of histone modifications through the use of
high-throughput sequencing protocols \citep{wang08combinatorial}. It
has been suggested that combinations of such histone modifications in
a given region, referred to as chromatin states, can be used as a mark
of the functional role of the region \citep{ernst11mapping}. One of
the proposed chromatin states, the ``5-enhancer'' (shortened to ``SE'' in
part of the following text), is suggested to correspond to regions that
play a role in gene regulation by providing accessible binding sites
to several transcription factors. It is thus interesting to see
whether different TFs indeed shows a higher than expected density of
experimentally determined binding events inside these regions. To
investigate this, we considered a collection of 82 tracks of
experimentally determined TF binding events in blood cells (cell type
gm12878) generated through the ENCODE project. The tracks are
originally of type Segments, corresponding to called signal peaks of
ChIP-seq experiments \citep{kim2005high}. These peak segments are
around 100~bps
long, reflecting experimental inaccuracy in the determination of
binding sites that are themselves around 5--25 bp long \citep{wingender1996}. The real
binding sites are often, but not always, located around the center of
these peak regions. In our analyses, we used the midpoints of the peak
regions as binding site locations. For each TF, we then tested whether
the binding locations occurred inside regions in the ``5-enhancer''
chromatin state more than expected by chance.

An analysis of the direct relation between TF binding locations and
chromatin states might be strongly confounded by a common relation to
gene locations. To reduce this potentially confounding factor, we
focused the study of the relation between TF binding and enhancer
states on only contiguous regions of size~$>100$~kb, that are more than
100~kbps away from the nearest gene.
Parts of these regions are located in centromeres, where neither TF
binding events nor chromatin states can be mapped. To avoid any bias
due to this, we constrained the analysis regions to only part of the
regions being located in the chromosome arms. There is a total of 580 such
regions in the human genome (using the Ensembl gene definition for
computing distance from genes), ranging in size from 100~kbp to 2.6 Mbp
and covering a total of 151 Mbps.

%f4 #&#
\begin{figure*}[t]

\includegraphics{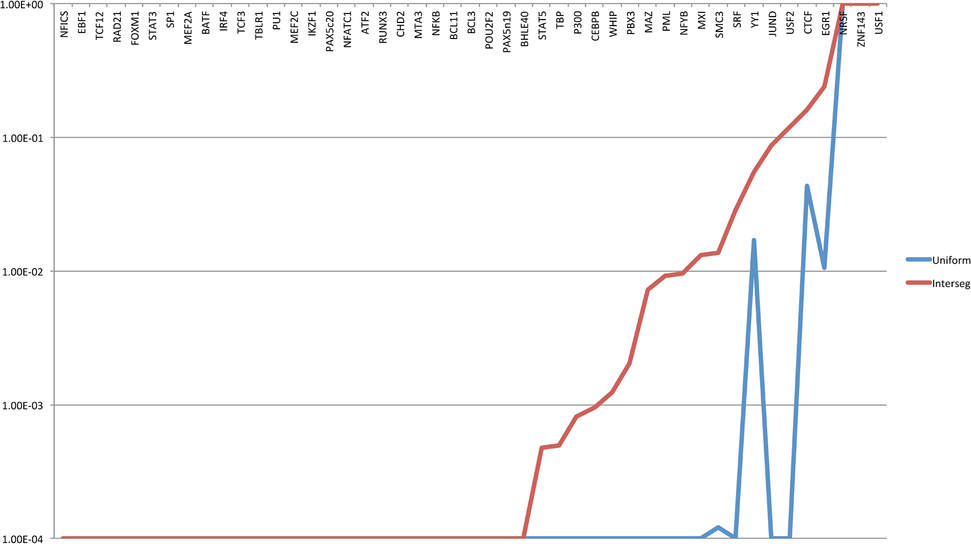}

\caption{$p$-values for hypothesis testing of whether midpoints of
ChIP-seq peaks for 47 different TF were
located more than expected inside regions of the chromatin state
5-Strong Enhancer. $p$-values were computed for two different null models:
random location of the midpoints or preserving the inter-point distances.
The TFs on the $x$-axis were sorted according to
the $p$-value achieved when preserving inter-point distances in the
null model.}\label{fig:PvalsAndKexclCentromers}
\end{figure*}

As can be seen from Figure~\ref{fig:Kboxplot}, ENCODE tracks display a
strong clustering tendency across different scales for a large number of
tracks of different types. The scaled Ripley $K$ values are described in
Section~\ref{sec:clustering}. All the collections show a typical
clustering tendency well beyond the neutral value of 1. Based on these results,
we claim that clustering is typical for genomic data of this type. We
observe very
few data sets where we find repulsion. Case B in Section~\ref{sec:ordering}
shows that clustering may give increasing $p$-values for null models:
if we
reduce or remove the clustering in the stochastic model, that is,
reduce the preservation,
then the $p$-values decrease. Hence, the $p$-value from the null models
are ordered according to increasing preservation of the clustering.
When testing the clustering it is important to apply a scale that is
adapted to the length of the observed property, for example, TFs. The
ordering of the $p$-values depends on the scale of clustering relative
to the
length of the properties (e.g., genes) in the
other tracks used in the test.

Furthermore, we tested whether ChIP-seq peaks for 47 different TFs transcription
factors were located more than expected inside regions of chromatin
state \mbox{5-Strong} Enhancer. $p$-values for two different null models with
random location of the CHIP-seq peaks or preserving the
inter-distances from the original tracks are shown in Figure~\ref{fig:PvalsAndKexclCentromers}. A total of 81 tracks of the TF
ChIP-seq peak region for the cell type gm12878 were retrieved from the
ENCODE data collection and analyzed against Strong Enhancer inside
regions of size~$>100$~kb that were more than 100~kbps away from the
nearest gene. For 34 of these tracks, there were less than a total of
20 peaks across all analysis regions, and they were removed from the
analysis. The $p$-values were computed based on Monte Carlo, using 10\mbox{,}000 samples, thus giving a minimum achievable $p$-value of 1E$-$4. For
some TFs, this minimum $p$-value was achieved using either null
model. For other TFs, either null model resulted in a $p$-value of 1. In
all cases where the two null models resulted in different $p$-values,
the null model that preserves inter-point distances gave the highest
$p$-value.

As we can see from Figure~\ref{fig:PvalsAndKexclCentromers},
very low $p$-values are reached for many
of the tests, confirming that the 5-Strong Enhancer chromatin state
captures histone modification patterns indicative of TF
binding. Indeed, when considering the union of binding locations
across all TFs, the relation between TFs and SE is highly significant
($p<0.00001$) for either null model. Our interpretation of the results
is that it clearly appears to be a relation of TF binding and the
5-Strong Enhancer chromatin state, but that the data limitation due to
only considering regions that meets the strict criteria above does not
allow a conclusion to be drawn regarding this relation for all TFs,
when considering only the behavior in these regions. The systematic
difference between $p$-values achieved using the two null models then
reflects that the null model preserving inter-point distances more
accurately portrays the possibility of concluding on the TF--SE
relation, while the null model disregarding the clustering of TF
points (inter-point distances) gives $p$-values that are lower than the
degree of certainty that can really be assigned to the TF--SE relation
in the considered analysis regions.

%s6 #&#
\section{False Rejections of Null Models Using Simulated Data}\label{sec:Falserejections}

In this section we perform hypothesis tests based on simulated
data with a clustering representative for genomic data. One
test has synthetic tracks for points and segments and another test
uses real TF tracks and simulated segment tracks. We generate the
tracks independently from each other, so the null hypothesis of
independence should not be rejected for any of the tests. In both
cases, we
get many false rejections if we assume uniform locations of points, but
good results when we preserve inter-point distances.

%t1 #&#
\begin{table*}[b]
\tabcolsep=0pt
\caption{Number of falsely rejected null hypotheses under different
combinations of data generation procedures and testing assumptions. Two
tracks of points and segments, respectively, are generated
independently, and then tested for significant relation. The different
columns correspond to whether points or segments are generated
uniformly (Poisson) or with a tendency for clustering. The different
rows correspond to whether points or segments are assumed to be random
in the null model, as well as whether location is assumed to be
uniformly distributed or according to a preserved empirical
distribution of inter-element distances}\label{tab:generationVsAssumption}
\begin{tabular*}{\textwidth}{@{\extracolsep{\fill}}lccc@{}}
\hline
&\multicolumn{3}{c}{\textbf{Generation}}\\[2pt]
\cline{2-4}
\rule{0pt}{12pt}\textbf{Assumption}             & \textbf{Uniform} & \textbf{Clustered points} & \textbf{Clustered segments} \\
\hline
Uniform point location (analytic) & 0/100   & 17/100           & 0/100              \\
Uniform point location (MC)       & 0/100   & 19/100           & 0/100              \\
Preserving inter-point distances  & 0/100   & 0/100            & 0/100              \\
Uniform segment location (MC)     & 0/100   & 0/100            & 0/100              \\
\hline
\end{tabular*}
\end{table*}

The previously presented genomic cases confirm that a null model with
a higher level of preservation typically gives higher $p$-values on
real data. However, they do not tell us which null model should be
preferred. As the simple null models will typically be easier to
implement, will often allow computationally fast analytical solutions
and will typically give more significance, they may be a tempting
choice for a practitioner. However, when their assumptions are not
met, there is a severe risk of false positive findings, due to the
failure of the null model to account for intrinsic characteristics of
the data, unrelated to the null and alternative hypotheses that are on
trial.

In order to study the potential severity of false positive findings
due to unrealistic null models, we performed a simulation study. Two
tracks were generated independently, but with various intrinsic
clustering-related properties. They were then tested for a relation
under different null models. The results are shown in
Table~\ref{tab:generationVsAssumption}. The synthetic tracks were
generated according to the approach described in
\citet{sandve2010}. Independent points were generated according to a
Poisson distribution with $\lambda=0.01$. Clustered points were
generated under an intra-cluster Poisson distribution with
$\lambda=0.1$ and inter-cluster Poisson with $\lambda=0.01$, with
each point having a probability $0.3$ of forming a new cluster.
Segments were generated similarly to points, with distance between
consecutive segments following a Poisson distribution with $\lambda= 0.01$.
Lengths of segments were distributed
uniformly between $10$ and $100$ base pairs. For each combination,
100 separate tests were performed and the number of false rejections
reported after multiple testing correction at 20\% FDR. We notice that
inappropriate assumptions could lead to up to 19\% of null hypotheses
being falsely rejected after correction for multiple testing.

Preserving more of the
individual properties (inter-element distances) was the safe choice,
essentially avoiding false rejections, while assuming uniform point
locations resulted in a high degree of false rejections, whether the
test was resolved analytically or by Monte Carlo simulation. For this
particular test, using a too simple assumption on segment location
(assuming uniform location for segments that were in reality
clustered) presented less of a problem. The reason for this is that
the autocorrelation between values of $X_i$, as discussed in Section~\ref{sec:ordering}, would be relatively low, and thus not lead to any
strong underestimation of
$p$-values.

It was shown in the previous two sections that using simple null
models led to lower $p$-values and more rejections when testing the
relation of TF binding to genes to certain chromatin
states. Although it would be tempting to consider the higher
significance as a sign of better power of the testing setup, the
assumption of uniform TF binding location is problematic and could
lead to $p$-values being underestimated. We have also shown on purely
simulated data that too simple and unrealistic assumptions can lead to
a high degree of false rejections. Here, we combine the real data of
TF binding with simulated segment data having the same characteristics
as genes and chromatin states, but where the simulated data is
generated independently from TF binding locations and chromatin states.
The null hypothesis
should then not be rejected in any test after multiple testing
correction.

For each chromosome band with at least 5 MitF binding sites, we tested
whether these binding sites occur differently than expected inside
simulated segments. This resulted in $H_0$ being rejected in 1 out of
73 bins at 10\% FDR when assuming uniform MitF locations. However,
when performing the tests only on 14 bins with a more satisfactory
amount of data (at least 10 MitF binding sites), the null hypothesis
is rejected in 4 out of 14 bins (still at 10\% FDR). This high rate of
false rejections suggests that part of the significance observed for
MitF versus genes or chromatin states under the assumption of uniform
location is likely due to underestimation of $p$-values due to the
inadequacy of this null model. Conversely, preserving the empirical
distribution of inter-MitF distances leads to no rejections of $H_0$
at 10\% FDR, either when testing in all 73 bins or in the 14 bins with
most MitF binding sites. This suggests that the preservation of
inter-point distances is able to capture the intrinsic structure of
the MitF track in an appropriate manner.

In summary, we find that the choice of null model strongly influences
the results. Mainly, the difference is that a null model preserving
more of the observed data yields higher $p$-values. Tests on simulated
data show that an overly simple null model, preserving too little of
the observed data, can lead to a large number of false rejections,
even after correcting for multiple testing.

%s7 #&#
\section{Discussion}\label{sec:discussion}

In this paper we have studied the choice of Monte Carlo null
models. We have defined the Monte Carlo state space as
the (finite) set of allowed resamplings of the observed data, and
defined a Monte Carlo null model preservation hierarchy.
We have discussed the null complexity principle, namely, that an
ordering of preservation may imply a corresponding
ordering of statistical significance (i.e., of estimated $p$-values),
and illustrated the use of
our result on real data sets of general interest.

The choice of null model is very application dependent, so it is
difficult to give general guidelines. However, two general approaches
are as follows: (1) to be conservative and choose the largest $p$-value
and (2) use the most restricted null model (which, however, should
still have sufficient freedom of variability to provide an efficient
test), so that we are ``close to the truth,'' that is, faithful to
restriction given by the phenomenon under study. Because of the null
complexity principle, approaches (1) and (2) will usually coincide.

A fundamental feature of the Monte Carlo approach to statistical
inference is that conclusions may only be drawn regarding the actual
observed data. In other words, there is no prospect for
generalizations to any (hypothetical or real) population. While some
may see this as a serious drawback of Monte Carlo methods, we feel
that this line of objection to randomization methodology is often
quite misguided. Obviously, the idea of random sampling from a
population is both useful and extremely entrenched in classical
statistics. However, often is it very hard to even conceive of the
``population'' in which random sampling is supposed to take
place. Genomics and DNA sequences are good examples of
this. In many cases, the Monte Carlo method is simply a more natural
approach: we do not wish to draw conclusions from a sample to a
population, it is really the (single, unique) sample itself that we
are genuinely interested in. In this paper we have focused on examples from
genetics since this is our main interest and motivation for the paper.
But similar problems are encountered in other areas such as ecology, as
documented in a separate
report; see \citet{ferkingstad2013}.

An interesting topic for future work would be to study the
implications for the multiple hypothesis testing setting. For a
discussion of some computational and conceptual challenges of Monte
Carlo multiple testing, see \citet{sandve2011}. The multiple testing
problem is particularly important in genomics, but it also appears in
ecology; see, for example, \citet{gotelli2010}.

Our main focus has been avoiding false positives due to too simple
null models. Of course, false negatives also occur, and the effect of
differing null models on the power of tests should be further studied.
In order to avoid underpowered tests, a very general advice is the
following. Most test statistics in the paper are based on counting,
hence, the variance of test statistics decreases as $1/N$ where $N$ is
the number of samples. But observations may be correlated, reducing
power. We may have very high correlation between a large number of
observations. It is important to be aware of this and try to find test
statistics where the correlation between observations is as small as possible.

Finally, we have also considered a third type of null model preservation,
where the data is a sequence of categorical variables, for example,
\ldots\break ACGT\ldots for a DNA sequence. The distribution for each
variable depends on the value of the previous $n$ variables. In this
model, it is possible to have the same probability distribution for
sequences of length $n$ as in the observed data.
Then, increasing $n$ implies preservering more of the
probabilistic structure of the original data.
We omitted this material to make the paper shorter and more focused. A
separate paper on this topic is in preparation.

\section*{Acknowledgments}
We thank Knut Liest{\o}l, Marit Holden and Arnoldo Frigessi, as well as
the editor, an associate editor and two anonymous referees, for very
helpful comments and suggestions.

%suskaldyti doi

% imsref loaded by aiste.veprauskaite, 2014-05-27 15:53:15
% imsref loaded by aiste.veprauskaite, 2014-05-27 16:01:42
% imsref loaded by aiste.veprauskaite, 2014-05-28 08:23:27
%

\end{document}